\documentclass[aps,prb,showpacs,twocolumn,groupedaddress]{revtex4-1}
\usepackage{times,xspace}
\usepackage{amsfonts}
\usepackage{amsfonts}
\usepackage{bm}
\usepackage{amssymb}
\usepackage[final]{graphicx}
\usepackage{color}

\begin{document}

\title{Vortex-induced dissipation in narrow current-biased thin-film superconducting strips}
\author{L.N. Bulaevskii, M.J. Graf, C.D. Batista, }
\affiliation{Theoretical Division, Los Alamos National Laboratory, Los Alamos, New Mexico 87545}
\author{V.G. Kogan}
\affiliation{Ames Laboratory DOE, Ames, Iowa 50011}

\date{\today}

\begin{abstract}
A vortex crossing a thin-film superconducting strip from one edge to the other, perpendicular to the  bias
current, is the dominant mechanism of dissipation for films of thickness $d$  on the order 
of the coherence length $\xi$ and of width $w$ much narrower than the Pearl length $\Lambda\gg w\gg\xi$.
At high bias currents,  $I^*<I<I_c$, the heat released by the crossing of a single vortex  suffices to 
create a belt-like normal-state region across the strip, resulting in a detectable voltage pulse.
Here $I_c$ is the critical current at which the energy barrier vanishes for a single vortex crossing.
The belt forms along the vortex path and causes a transition of the entire strip into 
the normal state. We estimate $I^*$ to be roughly $I_c/3$. 
Further, we argue that such ``hot'' vortex crossings are the origin of dark counts in 
photon detectors, which operate in the regime of metastable superconductivity at currents between $I^*$ and 
$I_c$. We estimate the rate of vortex crossings and compare it with recent experimental data for dark counts. 
For currents below $I^*$, i.e., in the stable superconducting but resistive regime, 
we estimate the amplitude and duration of voltage pulses induced by a single vortex crossing.
\end{abstract}
\maketitle

\section{Introduction}

Dissipation in  superconducting wires thinner 
than the  coherence length $\xi$ have been thoroughly studied 
 both theoretically \cite{Ambegaokar,McCumber} and experimentally. \cite{Tinkham}
In these one-dimensional (1D) superconductors the dissipation arises due to  $2\pi$-phase 
slips occurring in segments of length $\xi$ of a wire that becomes
temporarily normal. 
Langer and Ambegaokar\cite{Ambegaokar} treated
the problem of dissipation in 1D wires with ring geometry within the theory of nucleation rates 
of current-reducing fluctuations in a superconductor.
The transition between states with different currents in a ring occurs via the
nonstationary state  described by the saddle point solution of the
Ginzburg-Landau (GL) functional. Langer and Ambegaokar
\cite{Ambegaokar}  found such a solution and the corresponding free energy
difference or barrier,  ${\cal U}$, between the original metastable state
with current  and the saddle point state (see also
Ref.~\onlinecite{Ov}). 
Later McCumber and Halperin
derived the attempt frequency $\Omega$ in the phase-slip rate, $R=\Omega\exp(-{\cal U}/T)$,
using time-dependent GL theory.\cite{McCumber} 

The problem of dissipation in superconducting thin-film strips 
with the thickness $d$ much smaller than the London penetration depth $ \lambda$, and  of  
width $w$ much smaller than the Pearl length, $\Lambda=2\lambda^2/d\gg w$, has been  extensively discussed   in the
context of a possible Berezinsky-Kosterlitz-Thouless (BKT) transition
in  superconducting films.\cite{Goldman,Repaci,BKT} The interest in
current-carrying  thin-film strips has been revived recently in
search for   quantum tunneling of vortices,  
\cite{Ao,Stephen,Iengo,kogan1} their dynamic behavior,\cite{Kirtley}
and the observation  of so-called ``dark counts'' in
superconductor-based photon detectors. \cite{Kit,Engel}  The 
detector consists of a long and thin superconducting  strip
carrying currents slightly below the critical value.  
Typically, in NbN photon detectors $w$ is of
the order of 100 nm or more and $d \approx 4-6$ nm, 
while the zero-temperature coherence length $\xi(0)\approx 4$ nm. 
The low-temperature London penetration depth   $\lambda\approx 350\,$nm so 
that the Pearl length\cite{Mondal} $\Lambda\approx 40\,\mu$m $\gg w$.

When a photon interacts with the strip it   induces a hot spot in the film that drives a belt-like region across the strip in the normal state. 
Consequently, a voltage pulse 
caused by the current redistribution between the superconducting strip and a parallel shunt resistor is detected. 
After    the 
normal belt of the strip cools down, the strip returns to superconducting state.
Thus, single photons can be 
detected and counted by measuring voltage pulses. 
However,   similar pulses are recorded even without photons (dark counts). 
These voltage pulses have  peak amplitudes similar to photon-induced pulses.\cite{Kit1} 
Therefore, one can conclude that dark counts are also caused by nucleation of  normal   belts  across the strip. 
In both cases and in the absence of a shunt, the entire strip undergoes   transition  into the normal 
state   due to heat released by the bias 
current in the normal belt region. 

In fact, the observation of dark counts means that the superconducting strip, at bias currents slightly below 
the critical current, is in a metastable state. Photons or fluctuations   trigger the  transition from this 
state  to the normal state. Thus, the central question is what kind of fluctuations trigger the transition in the case of dark counts. 
The origin of dark counts is still debated (see Refs.~\onlinecite{Kit,Engel}). The problem of dark counts is
related to the basic question of dissipation in thin films and wires and is of technological relevance because  fluctuations resulting in the 
formation of normal belt across the strip limit the
ability of superconducting circuits to carry supercurrents, in general, and
the accuracy of photon detectors, in particular.
In  the literature, dark counts are treated either within the formal framework of 1D phase slips in
thin wires or within the picture of vortex-antivortex   unbinding 
near the BKT transition (see Refs.~\onlinecite{Kit,Engel}).
Vortices crossing the strip were employed to explain dc current-voltage characteristics of thin-film strips.\cite{kogan1,Gurevich}

In this paper we discuss three types of possible fluctuations in  superconducting 
strips which result in dissipation. Each one causes
 transition  to the normal state from the 
metastable superconducting state when currents are close to the critical value $I_c$:\\
(a) Spontaneous nucleation of a 
normal-state belt across the strip with $2\pi$-phase slip  as in thin wires.
\\
(b) Spontaneous nucleation of a single vortex near the edge of the strip and its motion across 
to the opposite edge accompanied by a voltage pulse. 
\\
(c) Spontaneous nucleation of vortex-antivortex pairs and their unbinding as they move 
across the strip to opposite edges due to the Lorentz force, as well as  the opposite process of nucleation of  vortices and antivortices at the opposite edges and their annihilation in the strip middle. 

The energy barrier for the nucleation of a temporary normal 
phase-slip belt is too high to be of importance  because the belt   volume $\gtrsim d\xi w$ is large. 
We will show that such a barrier remains large at any current in the superconducting state. 
Consequently, belt-like $2\pi$-phase slips appear with extremely low
probability.  
On the other hand, as  proposed in
Refs.~\onlinecite{kogan1,Gurevich}, thermally induced vortex 
crossings in current-carrying strips result in $2\pi$-phase changes along the strip just as in  the 1D scenario and  hence cause dissipation.
For the case of quantum tunneling this mechanism of dissipation was discussed in 
Refs.~\onlinecite{Ao,Stephen,Iengo,kogan1}.
The free energy barrier  for vortex crossing is much lower than for belt-like $2\pi$-phase slips, since the vortex core volume is $d\xi^2\ll d\xi w$.
The energy cost of creating a vortex and moving it over the barrier is 
$w/\xi$ times smaller than for creating a belt-like phase slip. 
An important point is that such a barrier for vortex crossing vanishes as the current approaches $I_c$,
whereas the barrier for the belt-like phase slip  remains nonzero at any current.
As to the vortex-antivortex process of the point (c), we  show in the following 
that the corresponding barrier   is twice as high as for the single vortex process.
 
We evaluate the amplitude of a voltage pulse and its duration assuming that the belt-like
area around  the vortex path remains superconducting. We call this process a  ``cold'' pulse.
This is not always the case, because  vortex motion excites quasiparticles along the vortex path 
and their   energies depending on the bias current  may  suffice for  creation of a normal-state belt across the strip.
This will result in redistribution of current from the  superconducting strip to the shunt 
with the accompanied voltage pulse much bigger than for  ``cold'' pulses. 
Such a ``hot''  pulse will be similar to the one induced by  photons.
In the following we will estimate at what minimum bias current $I^*$ a single
vortex crossing can trigger a ``hot'' voltage pulse and a corresponding dark count.

\begin{figure}[tb]
\includegraphics[width=8cm]{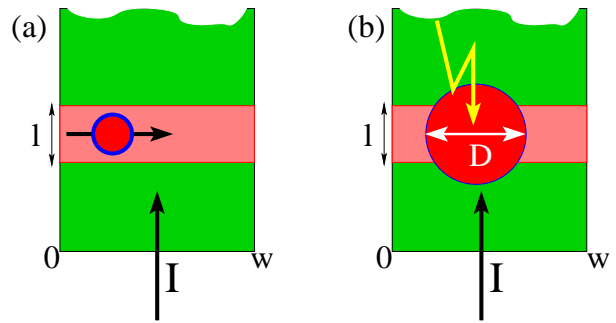}
\caption{(Color online)
Sketch of a segment of the strip in the presence of a bias current $I$. Panel  
(a):  a single vortex (blue circle) causes a ``hot'' crossing (pink belt). The width of the belt $\ell$  is of the order of 
superconducting coherence length. Panel
(b): a single photon creates a hotspot (red disc) and induces a subsequent ``hot'' vortex crossing (pink belt).
Both processes result in  detectable voltage pulses in a superconducting nanowire single photon detector (SNSPD).}
\label{fig0}
\end{figure}

Thus, we argue that dissipation and corresponding voltage pulses in strips are caused predominantly by vortex crossings. 
At high bias currents such crossings release  energy sufficient for the formation of a normal belt along the 
vortex trajectory, see Fig.~\ref{fig0}(a). Such a belt triggers the transition of the
whole strip into the normal state in the absence of a shunt resistor, as well as the redistribution of the bias 
current into the shunt in the case of  photon detectors. Note that a similar process happens when a  photon creates 
a normal ``hot'' spot on the strip. When this spot is sufficiently large, it destroys the superconducting path for the transport current 
  and the current redistribution leads to a voltage pulse,  the photon count. 
If  the hot spot does not disrupt completely the superconducting path, it will nevertheless lead to a decrease of the 
energy barrier for subsequent vortex crossings. At high bias currents, a ``hot'' vortex crossing can happen directly,
see Fig.~\ref{fig0}(a),
or through a hot spot area created by photon and forming a normal belt, which will result in signal detection, see Fig.~\ref{fig0}(b).

The layout of this paper is as follows: In Sec.~II we 
discuss three   energy barrier scenarios for vortex crossings. 
In Sec.~III we derive  dc current-voltage characteristics and evaluate the magnitude of 
induced voltage pulses.
The concept of ``cold'' and ``hot'' vortex crossings is introduced in Sec.~IV. 
In Sec.~V we compare our  results with data for dark count rates in NbN films.
\cite{Bartolf}
We summarize our results in Sec.~VI.

\section{Energy barriers and  vortex crossings}

In this section we derive  energy barriers for  three dissipative processes mentioned within   the 
GL  theory. Consider a thin-film strip of width $w\ll \Lambda$
and of length $L\gg w$. We choose the coordinates so that $0 \leq x \leq w$ and  $-L/2 \leq y \leq L/2$.  
Since we are interested in bias currents which may approach depairing values, 
the suppression of the superconducting order parameter must be taken into account.
We use the standard GL functional with respect to the  order parameter $\Psi({\bm r})$ (normalized to its zero-field value in the absence of current) and  the vector potential ${\bm A}$:
\begin{eqnarray}
{\cal F}[\Psi({\bm r}), \bm{A}]&=& \frac{H_c^2 d}{ 4\pi} 
 \int d{\bm r}\Big[-|\Psi|^2+\frac{1}{2}|\Psi|^4 \nonumber\\
 &+&\xi^2\left|\left(\nabla+i\frac{2\pi}{\Phi_0}{\bm A}\right)\Psi\right|^2
 + \frac{B^2}{8\pi}\Big]. 
 \label{GL}
 \end{eqnarray}

Here  $\Phi_0$ is the flux quantum, 
${\bm r}=(x,y)$ is a point on the film, $\bm \nabla $ is the 2D gradient, 
and $H_c = \Phi_0/ 2\sqrt{2}\pi\lambda\xi $ is the  thermodynamic critical field. 
The order parameter in the presence of a uniform bias current $I$ in zero applied magnetic fields and with no  vortices present, can be found by minimizing the GL functional and disregarding the current self-field, as is done, e.g., in Ref.\,\onlinecite{McCumber}. 
As discussed in the next section, this is an accurate approximation for $w\ll\Lambda$. Thus we obtain the solution: 
\begin{eqnarray}
&&\Psi_{\kappa}({\bm r})=(1-\kappa^2 )^{1/2} e^{-i\kappa y/\xi
+i\varphi_0},
\label{orderp1}\\ 
&&I=\frac{2w}{\pi\xi}I_0\kappa (1-\kappa^2 ), \ \ \ \ 
I_0=\frac{ c\Phi_0}{8\pi\Lambda}, 
\label{cur}
\end{eqnarray} 
where  $\varphi_0$ is an arbitrary constant phase.
The parameter $\kappa$ is proportional to the phase gradient and describes the order parameter suppression due to bias current. As a function of $\kappa$ the   bias current in the superconducting state is 
limited to the depairing current $I_{\rm max}=I_0(4/3\pi\sqrt{3})(w/\xi)$, 
corresponding to $\kappa_{\rm max}=1/\sqrt{3}$, as for the case of the 1D wire.

\subsection{Phase slip in the normal belt}

When dealing with the situation of fixed uniform current $I$ instead of 
vector potential $\bm{A}$ it is more suitable to work with the Gibbs free energy functional rather than 
the free energy functional, Eq.~(\ref{GL}). We perform the usual Legendre transform 
(see Ref. ~\onlinecite{Tinkham}) to obtain the corresponding free energy density:
\begin{equation}
f_I\{\Psi\}=
\frac{H_c^2}{4\pi}\left[-|\Psi|^2+\frac{1}{2}|\Psi|^4+\left(\frac{I\pi\xi}{2wI_0}\right)^2
|\Psi|^{-2}\right]. 
\label{f(I)}
\end{equation}
The equilibrium Gibbs free energy density for a given current is obtained by minimization with respect to $\Psi$. 
It jumps at the maximum current $I_{\rm max}$ from $f_I(I_{\rm max})=-(2/9)(H_c^2/8\pi)$ to zero at $I_{\rm max}$
as expected  for a first order transition.
Hence, the free energy barrier ${\cal U}$ 
for creation of a belt-like normal-state area with volume $V=\ell wd$
($\ell$ is the width of the belt along the $y$-axis) decreases from $(H_c^2/8\pi) V$ to 
$(2/9)(H_c^2/8\pi) V$ as the bias current increases from 0 to $I_{\rm max}$. 
The barrier never vanishes 
in this interval (``overheating'' with respect to bias current is absent). 
Note that for $w\gg\xi$ and $\ell\gtrsim \xi$ the barrier remains  very high in comparison with the temperature 
at all bias currents $I<I_c$ resulting in 
low probability for phase slips, except for temperatures close to $T_c$, where the barrier  vanishes as $(1-T/T_c)^2$.
 
\subsection{Single vortex crossing}

A vortex crossing from one strip edge to the opposite one induces
a phase slip without creating a normal region across the strip width. 
We will treat the vortex as a particle moving in the energy 
potential formed by the superconducting currents around vortex center inside the strip and by the 
Lorentz force induced by the bias current. We will derive the energy potential and find the vortex crossings 
rate (phase slips and corresponding voltage pulses) 
in the framework of Langevin equation for viscous vortex motion and invoke the known solution of the corresponding Fokker-Planck equation. 

In the presence of a vortex,   the order parameter in the current-carrying strip,  
disregarding its  suppression in the vortex core,  reads: 
 \begin{eqnarray}
&&\Psi({\bm r},{\bm r}_v)=\mu\exp\{{i[\varphi({\bm r},{\bm r}_v)-\kappa y/\xi+\varphi_0]}\}, \label{opv} \\
&&\mu^2=1-\kappa^2.
\end{eqnarray}
In this approximation, the vortex affects mainly the phase $\varphi({\bm r},{\bm r}_v)$ of the order parameter.  
To describe voltage pulses we need to know how the phase changes when the vortex moves across the strip. 
For simplicity we consider a vortex at   $ (x_v, y_v)=0$. 
As we ignore the change of the order parameter amplitude in  the vortex core,  
the current distribution is governed by the London equation (integrated over the film thickness):
\begin{equation}
h_z+ 2\pi(\Lambda/c) \,{\rm curl}_z {\bm g}=\Phi_0\,\delta ({\bm r}-{\bm r}_v ) ,
\end{equation}
where $\bm g$ is the sheet current density.  

For narrow strips, $w \ll \Lambda$, 
the field is approximately $h_z\sim   g/c$, whereas the term with derivatives is of
the order $\Lambda  g/cw$. Hence, in this limit,   supercurrents can be found by neglecting $h_z$ and the
corresponding vector potential  of the order $w/\Lambda$. \cite{kogan2,sasha}  
Introducing the scalar stream function $G({\bm r})$ such that 
 \begin{eqnarray}
{\bm g} ={\rm curl} (G \,\hat{ \bm z }) \,  
\label{gG}
 \end{eqnarray}
we reduce 
the problem   to
solving the Poisson equation:  
\begin{equation}
\nabla^2G= -(c\Phi_0/2\pi\Lambda)\delta({\bm r}-{\bm r}_v) \,.
\end{equation} 

Since the boundary condition at the strip edges requires
vanishing normal  components of the current, we have 
$G=0$ at $x=0,w$. Therefore, the problem is 
equivalent  to one in 2D electrostatics:   a linear charge at ${\bm r}_v$ between two parallel grounded plates  at $x=0,w$ with the known solution:\cite{Morse}
\begin{eqnarray}
G({\bm r})=\frac{I_0\mu^2}{\pi}\ln\frac{\cosh Y-\cos(X+X_v)}{\cosh Y-\cos (X-X_v)}\,,
\nonumber\end{eqnarray}
where   capitals   stand for coordinates in units of $ w/ \pi $, i.e., $x=X\, w/\pi$, $y=Y \, w/\pi$.

The  energy of a vortex at $x=x_v$ and $y=0$ is:
\begin{equation}
\epsilon_v= \frac{\Phi_0}{2c} \,G(x_v,0) ,
\end{equation} 
with the standard cutoff $\xi$ at the vortex core.\cite{kogan2}  In the presence of a
uniform bias current the energy barrier reads:
\begin{eqnarray}
&&{\cal U}(X_v)=\mu^2\epsilon_0\left[\ln\left(\frac{2w}{\pi\xi}\sin X_v\right)
-\frac{I}{\mu^2I_0}X_v\right],\quad
\label{potential} 
\\
&& \epsilon_0=\frac{\Phi_0^2}{8\pi^2\Lambda}=\frac{H_c^2}{8\pi}(4\pi\xi^2),
\label{vortex_energy} 
\end{eqnarray}
where  $\epsilon_0$ is the characteristic energy of a vortex 
in thin films. The vortex energy ${\cal U}(X_v)$ is maximum at 
$ X_s=\tan^{-1}( \mu^2I_0/I)$ and   the energy barrier is given by
\begin{equation}
\frac{{\cal U}}{\mu^2\epsilon_0}= -\frac{1}{2}\ln\left[\frac{\pi^2\xi^2}{4w^2}\left(1+\frac{I^2}{\mu^4I_0^2}\right)\right]-\frac{I}{\mu^2I_0}
\tan^{-1}\frac{\mu^2I_0}{I} .
\label{eq14}
\end{equation}
This barrier decreases with increasing current and turns zero at a critical value on the order of the depairing GL current: 
\begin{equation}
I_c=  \frac{2\mu_c^2wI_0}{\pi e\xi} =  \frac{c\Phi_0\mu_c^2 }{8\pi^2e\lambda^2  \xi}\,wd ,
\label{Ic}
\end{equation}
here $e=2.718$. One can see that the critical current $I_c$ is slightly smaller than $I_{\rm max}$ discussed above.

Since the vortex mass is negligibly small, we use the equation of purely diffusive motion (only includes first order time
derivative) for describing the vortex propagation between $x=0$ and $x=w$:
\label{eq}\begin{equation}
{\gamma}\frac{dX_v}{dt}=-\frac{d{\cal U}(X_v)}{dX_v}+F(t),
\label{eom}
\end{equation}
where $\gamma={w^2\eta}/{\pi^2}$ and 
\begin{equation}
\eta=\frac{\Phi_0^2 }{2\pi\xi^2c^2R_{\square}}\,,
\label{motion}
\end{equation}
 is the Bardeen-Stephen drag coefficient for film with 
 $R_{\square}=\rho_n/d$ being the film's sheet resistance slightly above $T_c$.  
 $F(t)$ is the 
Langevin random force obeying statistical averages $\langle F(t)\rangle=0$ and
$\langle F(t)F(t')\rangle=2\gamma T\delta(t-t')$.

The vortex motion described by Eq.~(\ref{eom}), takes place in the interval $a<x<w-a$, where $a$ is of the order of $\xi$
(the energy of the system cannot be described by the potential (\ref{potential}) in the intervals $w-a<x<w$ and $0<x<a$).
The most crucial interval for vortex motion is near the point $x_s=X_s w/\pi$, where the vortex should overcome the 
potential barrier. Thus $x_s$ should be inside the interval $(a,w-a)$, i.e., the  conditions $\xi\ll w$ and  $I<(e/2)I_c$ should be 
fulfilled to consider the motion of vortex in the interval $0<x<w$.
To compute the average velocity in the interval $0<x<w$, we consider the diffusion problem of a single particle
that propagates in the interval $-\infty <  x < \infty$ under the effect of the periodic potential $\epsilon_{v}(x)=\epsilon_{v}(x+w)$ and the Lorentz force.
The average velocity is obtained from the known stationary solution for this periodic model 
(see Ref.~\onlinecite{Risken}). This approach was previously 
used by Gurevich and Viinokur \cite{Gurevich}.

The corresponding Fokker-Planck equation (Smoluchowski equation)
for the probability current in the case of the periodic potential 
has a stationary solution with the 
statistical average vortex velocity $\overline{v}$ given by   \cite{Risken}
\begin{eqnarray}
&&\gamma\overline{v}=
\frac{\pi T {P}}{Z_+(\pi)Z_-(\pi)-{P}\int_0^\pi dx e^{-{\cal U}(x)/T}
Z_+(x)}, \\
&&Z_{\pm}(x)=\int_0^x du \,e^{\pm {\cal U}(u)/T}, \quad
{P} = 1-e^{-\pi p}
 \label{meq}
\end{eqnarray}
where $\overline{v}\equiv \overline{\dot{X}}$ and $p=\nu I/\mu^2I_0$. Except for temperatures close to $T_c$ the parameter
 $\nu=\mu^2\epsilon_0/T\gg 1$ . 
At large $\nu$ the function $\exp[{\cal U}(x)/T]$ has a sharp maximum between $0$ and $w$, 
while the function $\exp[-{\cal U}(x)/T]$ has two sharp 
maxima at the edges of this interval. 
Since the integral $Z_+(\pi)$ has the analytic solution\cite{Integrals} 
\begin{equation}
 \int_0^{\pi} dx\,e^{-px}\sin^{\nu}x = \frac{\pi\exp(-\pi p/2)\Gamma(\nu+1)}{2^{\nu}|\Gamma(1+\nu/2+ip/2)|^2} ,
\end{equation}
where $\Gamma(x)$ is the Gamma-function and $\nu>-1$, we obtain the asymptotic solution for $\nu\gg 1 $:
\begin{equation}
Z_+(\pi) \approx
\left(\frac{2w}{\pi\xi}\right)^{\nu} \sqrt{\frac{2\pi}{\nu}} 
\left(1+\frac{p^2}{\nu^2}\right)^{-\frac{\nu+1}{2}}
e^{-p\tan^{-1}( \nu/p )}.
\end{equation}
Evaluating $Z_-(\pi)$ we note that the main contribution comes from the 
regions near the edges, where we approximate $\sin(x)=\sin(\pi-x)\approx x$ and replace 
the low integration limit  by $\pi\xi/w$ and the upper one by $\pi-\pi\xi/w$. We 
obtain the asymptotic limit
\begin{equation}
Z_-(\pi) \approx \left(\frac{2w}{\pi\xi}\right)^{-\nu}\left(\frac{w}{\pi\xi}\right)^{\nu-1}
\frac{e^{\pi p}+1}{\nu-1}.
\end{equation}
In the integral $\int_0^\pi dx e^{-{\cal U}(x)/T}Z_+(x)$, the function $Z_+(x)$ 
reaches  maximum at $x=\pi$ and
is small at low $x$. Hence, the main contribution to this integral 
comes from the region near $x=\pi$:  
\begin{equation}
\int_0^\pi dx\, e^{-{\cal U}(x)/T}Z_+(x)
\approx\left(\frac{2w}{\pi\xi}\right)^{-\nu}
\!\!
\left(\frac{w}{\pi\xi}\right)^{\nu-1}
\!\!
\frac{e^{\pi p}}{\nu-1}
 Z_+(\pi).
\end{equation}
It then follows that the dependence of the average vortex velocity $\overline{v}$ on   $I$
at large $p$ and $\nu$  is given by   
\begin{eqnarray}
 \gamma \overline{v} \approx T\left(\frac{\pi\nu^3}{2}\right)^{1/2}\left(\frac{\pi\xi}{w}\right)^{\nu-1}
Y\left(\frac{I }{\mu^2I_0}\right) \label{single1},\\
 Y(z)=(1+z^2)^{(\nu+1)/2}\exp[\nu z\tan^{-1}(1/z)]\label{defY}.
\end{eqnarray}
Note the strong power-law dependence of $ \overline{v}$ on the strip width $w$.
 
For large currents, $I\gg I_0$, this expression reduces to
\begin{eqnarray}
 \gamma \overline{v}\approx
T\left(\frac{\pi\nu^3}{2}\right)^{1/2}\left(\frac{w}{\pi\xi}\right)^2\left(\frac{I}{I_c}\right)^{\nu-1},
\label{powerlaw}\end{eqnarray}
with $I_c$ given by Eq.~(\ref{Ic}). 
Note that the average velocity changes drastically near the critical current $I_c$, 
where the energy barrier vanishes. 
Such defined critical current is about 16\% smaller 
than the standard depairing current $I_{\rm max}$ defined for 1D wires (vanishing energy barrier for phase slips 
in wires, see Ref.~\onlinecite{McCumber}).\cite{footnote1}

In the case of multiple simultaneous vortex crossings happening in different parts of the strip, we must account for their interactions. The interaction   of
vortices situated at $(X_1,0)$ and  $(X_2,Y)$ has been evaluated   in Ref.\,\onlinecite{kogan2}:  
\begin{equation}
 \epsilon_{\rm int} = \epsilon_0   \ln  \frac{\cosh Y -\cos(X_1+X_2)} { \cosh Y  -\cos  (X_1-X_2) } 
 .\qquad
\label{inter}\end{equation}
If vortices are separated by $y>w$ along the strip, the interaction is exponentially weak and   their crossings are uncorrelated. 
Accounting for both vortex and antivortex crossings (which are equivalent by symmetry), we estimate  the rate for multiple vortex crossings at $I<I_c$ as 
$R\approx(2L/\pi w)\overline{v}$. 

Finally, we obtain the asymptotic estimate for the rate:
\begin{equation}
R \approx \frac{4Tc^2R_{\square}L}{\Phi_0^2w }\left(\frac{\pi\nu^3}{2}\right)^{1/2}\left(\frac{\pi\xi}{w}\right)
^{\nu+1}Y\left(\frac{I }{\mu^2I_0}\right) .
\label{rate}\end{equation}
In obtaining this result we disregarded vortices crossing in the direction opposite to the Lorentz force, the corresponding probability for such processes is $\propto e^{-2 p} \ll 1$.
 We note that Gurevich and Vinokur took $L/\xi$ as the number of statistically 
independent vortex  crossings.\cite{Gurevich}  It differs by a factor $\xi/w\ll 1$ from our estimated number $L/w$ of independent crossings. Therefore, 
Ref.~\onlinecite{Gurevich} overestimates the rate. 
 
\subsection{Vortex-antivortex pair scenario}  

The energy of a vortex-antivortex pair (vortex-antivortex interaction included) was derived in
Ref.~\onlinecite{kogan2} and is
\begin{equation}
\frac{\epsilon_p}{\mu^2\epsilon_0} =   \ln\left[\frac{4W^2}{\pi^2 \xi^2} \sin X_1 \sin
X_2\,\frac{\cosh Y -\cos(X_1-X_2)} { \cosh Y  -\cos  (X_1+X_2) } 
\right] .
\label{inter2}\end{equation}
This energy increases with increasing separation $Y$, so that one expects the
lowest barriers for $Y=0$:
\begin{equation}
\frac{\epsilon_p}{\mu^2\epsilon_0} =\ln\left[\frac{4w^2}{\pi^2\xi^2}\sin X_1 \sin X_2 \,\frac{\sin^2[(X_1-X_2)/2]}
{\sin^2[(X_1+X_2)/2]}\right].
\end{equation}
One can show that if a pair is formed at $X_0$ and the pair
members are pushed apart a distance $2b$,   the
lowest energy increase (for a given $b$) corresponds to the initial
position $X_0=\pi/2$ in the middle of the strip. The energy barrier for
such a pair, in the presence of bias current $I$, is obtained by setting
$X_{1,2}=\pi/2\mp b$ and adding the Lorentz force contribution:
\begin{equation}
{\cal U}_p(b)=2\mu^2\epsilon_0\left(\ln \frac{ w \sin( 2b)}{\pi \xi}
-\frac{Ib}{\mu^2I_0}\right).
\end{equation}
This energy is maximum if $2b=\tan^{-1}(2\mu^2I_0/I)$ so that the
  energy barrier for vortex-antivortex pairs is given by
\begin{equation}
\frac{{\cal U}_p}{ \mu^2 \epsilon_0}= -  \ln\left[\frac{\pi^2\xi^2}{w^2}\left(1+\frac{I^2}{4\mu^2I_0^2}\right)\right]-
\frac{I}{ \mu^2I_0}\tan^{-1}\frac{2\mu^2 I_0}{I} .
\label{DF}
\end{equation}
For $I\gg I_0$ this barrier is twice as large than that for a single vortex 
crossing, Eq.\,(\ref{eq14}), and the ratio of these barriers increases for smaller currents. 
Note also that the core
contribution to the pair energy (neglected here) is at  least twice that for a single
vortex. 

Based on our estimates for the three different fluctuation scenarios presented here,  
we conclude that single vortex crossings are the main source for dark counts.

\section{Voltage induced by vortex crossing}

Let us now  find how the phase of the order parameter varies when a 
vortex crosses the strip. The current is expressed either in 
terms of the gauge invariant phase $\varphi$ or via the stream function 
$G$:  
${\bm g}=-(c\Phi_0/4\pi^2\Lambda)\nabla \varphi 
={\rm curl}[\,G \hat{{\bm z}}]$.  Written in components, this 
gives the Cauchy-Riemann relations for  functions
$[4\pi^2\Lambda_0/c\Phi_0\mu^2]G({\bm r})$ and $\varphi({\bm r})$. Hence they 
are real and imaginary parts of an analytic function of complex argument
$z=x+iy$:\cite{Morse} 
\begin{equation}
{\cal G}(Z)= \ln\frac{\sin [( X_v+Z)/2] }{\sin[ (X_v-Z)/2] }\, 
\label{Gphase}
\end{equation}
(recall: the capitals are coordinates in units of $w/\pi$, so that $0<X<\pi$, etc.). We then obtain
\begin{eqnarray}
\varphi({\bm r},{\bm r}_v)&=& {\rm Im}[{\cal G}(Z)]
\\
&=&\tan^{-1} \frac{\sin X_v \sinh (Y-Y_v)}{\cos X-\cosh (Y-Y_v)\cos X_v } .
\nonumber
\label{phase}
\end{eqnarray}
Note that  the characteristic length of variations  for $\varphi$ in both $x$ and $y$ directions is $w$. 
For long strips  of interest, $L\gg w$, and for distances $|Y-Y_v|\gg 1$, we have at the strip ends  $\varphi(\pm L/2)=\mp X_v$. Hence, when the vortex
moves from the strip edge    at $X_v=0$ to the opposite edge at $X_v=\pi$ and $|L/2-Y_v|\gg 1$, the phase
difference at the ends of the strip changes by $\varphi(L/2)-\varphi(-L/2) = 2 X_v=2\pi$, i.e., 
a vortex crossing results in a global phase slip of $2\pi$.

\subsection{DC voltage}

The motion of vortices   causes   the phase difference at the strip ends  to vary in time.  
Using the Josephson relation for the phase, we obtain  the induced voltage due to a single vortex crossing
\begin{equation}
V(t) =\frac{\Phi_0}{2\pi c}\,\frac{d}{d t} 
[\varphi(L/2)-\varphi(-L/2) ]= \frac{\Phi_0v(t)}{c w},
\label{V-v}
\end{equation}
where the vortex velocity is $v(t)=dx_v/dt=(w/\pi) d X_v/d t$ and we used $\varphi(L/2)-\varphi(-L/2)=2X_v$. A quasistatic approach employed here is justified as long as the characteristic crossing time  $\Delta t = w/v$ is large compared to   $L/c$.  
Note that for each crossing, i.e., for each voltage pulse between time $t$ and $t+\Delta t$ the relation
\begin{equation}
\int_t^{t+\Delta t} dt' \, V(t')=\frac{\Phi_0}{c}
\label{J-Phase}
\end{equation}
is satisfied as in the case of voltage pulses due to phase slips in 1D wires.\cite{McCumber}
Thus we obtain the average (dc) voltage 
 \begin{equation}
V_{\rm dc}  = \frac{\Phi_0}{c} R.
\label{v11}
\end{equation}
This relation also follows directly from comparing the dissipated  
power $V_{\rm dc} I$ with the work per unit time done by the Lorentz  force, $(\Phi_0 I/cw) w R$.
It is worth to remember that we have derived the  crossing rate assuming an isothermal strip. 
In continuous measurements of current-voltage characteristics   
at currents of the order of the critical one,  the strip temperature   is certainly higher than that of the bath. 
In principle, this heating may be reduced  using short bias current pulses. 

\subsection{Voltage pulses}

In this section we consider the time evolution of the voltage pulse $V(t)$ induced by  single vortex crossing.
Here we use the equation of vortex motion, Eq.~(\ref{eom}), 
for $X>X_s$ and neglect random forces (thermal noise).
Therefore the velocity is
 \begin{equation}
v\equiv {\dot x_v}=\frac{\pi\epsilon_0}{\eta w}\left( \frac{I}{I_0} -\mu^2 \cot X\right) . 
\label{v}
\end{equation}
This can be written in the form
 \begin{equation}
 {\dot X}= \beta\left( \cot X_s -  \cot X\right) ,\quad  \beta =  \frac{\pi^2\epsilon_0\mu^2}{\eta w^2},
\label{v1}
\end{equation}
which is valid for $X>X_s$. 
It is worth noting that for currents of the order of $I_c$ the saddle point is very close 
to the strip edge,
\begin{equation}
X_s\approx \frac{I_0 \mu^2}{I}=\frac{e\pi}{2}\, \frac{\mu^2I_c}{\mu_c^2 I}\,\frac{\xi}{w}  \ll 1\,.
\label{X_small} 
\end{equation}
Integration of Eq.~(\ref{v1}) results in an implicit
 solution for $X(t)$:
 \begin{equation}
    X(t)\cos X_s+\sin X_s\ln\sin\left[ X(t)-  X_s\right] =
 \frac{\beta\left( t-t_0\right)}{ \sin X_s}. 
 \quad  
\label{t(X)}
\end{equation}
We  choose the constant $t_0$ so that   $t=0$ corresponds to the vortex exit   at $X=\pi$.
Note that any instant for which $0<X(t)< X_s$ is beyond this approximation, 
because in this early time interval the process is described by 
thermal activation rather than by the equation of motion (\ref{v}) with random force omitted. 
The   instant for which $ X(t)= X_s$  is also inappropriate as an initial moment, because at this point 
the velocity vanishes, $\dot X=0$. Thus Eq.~(\ref{t(X)}) can be written as
 \begin{eqnarray}
 [X(t)-\pi]\cos X_s+\sin X_s\ln\frac{\sin\left[  X(t)-  X_s\right]}{\sin X_s}
= \frac{\beta\, t}{\sin X_s}  . 
\nonumber\\
\label{t(X)1}
\end{eqnarray}
Clearly, $X(0)=\pi$ and $ X(t\to-\infty) = X_s$.  
Hence, formally, the motion from the saddle point $X_s$ to the edge takes 
infinite time because the velocity goes to zero as $X\to X_s$. 
In reality, the dynamic viscous vortex motion starts at some distance from the saddle 
point where the vortex is kicked by random force (an activation driven  process) and the 
total "time-of-flight" is finite. To see this, consider the situation of large currents for 
which $X_s $ is given by Eq.~(\ref{X_small}) and 
\begin{eqnarray}
  X(t)-\pi +  X_s\ln\frac{\sin\left[  X(t)-  X_s\right]}{ \sin X_s}  = \frac{\beta\, t}{  X_s}  . \quad
\label{t(X)2}
\end{eqnarray}
Denote as $\delta X$ a small distance from the saddle at $X_s$ and 
evaluate the time $\tau_0$ of motion from $X_s+\delta X$ to the edge $X=\pi$:
 \begin{eqnarray}
- \frac{\beta\,\tau_0}{  X_s} =X_s+ \delta X  -\pi +  X_s\ln\frac{   \delta X }{  X_s}  . \quad
\nonumber
\end{eqnarray}
Since both $\delta X$ and $  X_s$ are small, 
all terms on the right hand side, except for $\pi$, are negligible and we obtain
 \begin{eqnarray}
\tau_0\approx  \frac { \pi X_s}{\beta} = \frac {c\eta w^2}{\Phi_0 I}=\frac{w^2\Phi_0}{2\pi\xi^2cR_{\square}I}
\,,  
\label{T}
\end{eqnarray}
so that the time-of-flight $\tau_0$ does not depend on a particular choice of $\delta X$. In fact, this estimate coincides with the time it takes a vortex to cross the strip being pushed solely by the Lorentz force.

\begin{figure}[tb]
\includegraphics[width=8cm]{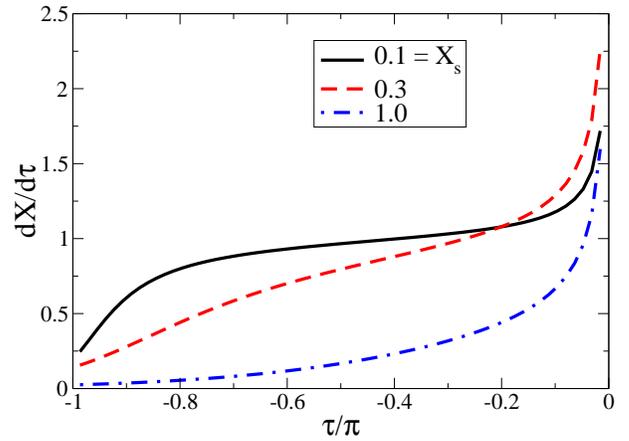}
\caption{(Color online)
The dimensionless vortex velocity $dX/d\tau$  versus 
$\tau$ for parameters $X_s=0.1, 0.3$, and 1.0. 
Note that $dX/d\tau=v/{\bar v}$, where $v$ is the velocity in common units and $ {\bar v}$ is the average velocity, which is identical to the one solely due to the Lorentz force.}
\label{fig1}
\end{figure}

Solving numerically  Eq.~(\ref{t(X)1}) for $X(t)$ and 
substituting the result in Eq.~(\ref{v1}) we obtain $v(t)$. The 
result is shown in Fig.~\ref{fig1}. For convenience, we use    
$ X_s/\beta$ as the unit of time. The 
dimensionless time $\tau=\beta t/ X_s$ varies between $-\pi<\tau<0$.

The divergence at the edge $x=w$ must be cut off  at distances of the order of
$\xi$ from the edge. We obtain from Eq.~(\ref{v1}) an estimate for the maximum 
velocity at the exit,
\begin{eqnarray}
v_{max} \approx  \frac{\phi_0 }{cw\eta}\left(I+\frac{e\,I_c}{2}\right) \,,
\label{v-max1}
\end{eqnarray}
where the critical current is given by Eq.~(\ref{Ic}). 

For large currents, $X_s\ll 1$, we solve Eq.~(\ref{t(X)2}) 
perturbatively:  $X=X_1+\delta X $ with $X_1=\pi+\beta t/X_s$ 
and $ \delta X\ll X_1$:
\begin{eqnarray}
X =\pi+  \tau - X_s\ln\frac{\sin\left(  X_s-\tau   \right)}{ \sin X_s} .
\label{X}
\end{eqnarray}
Thus the velocity for $X_s\ll 1$  is
 \begin{eqnarray}
\frac{d X}{d\tau}= 1 +X_s \cot (X_s-\tau)  \,,
\label{v3}
\end{eqnarray}
 the unity   corresponds to a constant velocity 
due to the Lorentz force, whereas the second term is caused by 
the vortex potential.

The velocity $v(t)$ is  peaked near the edge $x=w$ and it is of interest to estimate the width $\Delta \tau$
of this peak in the velocity and in the voltage $V(t)\propto v(t)$. 
The width $\Delta \tau$ is definition dependent.  For example, one can define  
it as the time interval between instants when $v=v_{\rm max}$ and time $\tau_1$ when 
$v=(v_{\rm max}+{\overline v})/2$, where ${\overline v}$ is the background velocity due to the current $I$. 
In dimensionless units, ${\overline v}$ corresponds to $dX/d\tau =1$. Thus we obtain 
\begin{eqnarray}
\tau_1\approx - X_s\,\frac{6wX_s+\pi\xi }{3wX_s-\pi\xi }   \,. 
\label{tau1}
\end{eqnarray}
with 
 \begin{eqnarray}
 \frac{ \pi\xi }{ wX_s } =\frac{2\mu_c^2 I}{e\mu^2I_c} < 1  \,, 
\label{inequal}
\end{eqnarray}
so that $\tau_1 <0$.  Since $|\tau_1|\sim X_s\gg\tau_m$, we estimate the width of the velocity peak near the edge as $\Delta t\sim \Delta\tau X_s/\beta \sim X_s^2/\beta$, where the fraction of order unity in Eq.~(\ref{tau1}) has been neglected. Therefore, the ratio of this width relative to the total crossing 
time $\tau_0$ of Eq.~(\ref{T}) is 
\begin{eqnarray}
\frac{\Delta t}{\tau_0} \approx \frac{  X_s }{ \pi  } \ll 1  \,.
\label{ratio}
\end{eqnarray}

\section{``Cold'' and ``hot'' vortex crossings}

A vortex moving from the saddle point $x=x_s$ to the strip edge $x=w$, during the time 
$\tau_0= w/{\overline v}$, excites quasiparticles along its path by the mechanism described by 
Larkin and Ovchinnikov.\cite{LO, Schmid} 
This mechanism is appropriate for dirty superconductors (for 
clean and intermediate clean regimes see Refs.~\onlinecite{Eschrig, Kopnin}).
Since NbN films are inherently   dirty, we can safely disregard the latter mechanism.
We estimate the total  energy transferred to  quasiparticles during the time $\tau_0$ along the vortex path as
\begin{equation}
Q\approx(\Phi_0I/c)\approx \frac{8\pi}{e}~\frac{H_c^2}{8\pi}~\frac{I}{I_c}w\xi d.
\end{equation}
This is, in fact, the work done by the Lorentz force on the vortex path of the length $w-x_s$. 
This energy is distributed near uniformly along the path at currents 
close to the critical current, 
because  the vortex velocity varies weakly for most of the crossing, see Fig.~1. In a belt of width $\ell$ along the $y$-axis 
 with the volume $V_b=\ell wd$, the energy increase  per unit volume is 
$(8\pi\xi/e\ell)(H_c^2/8\pi)(I/I_c)$. 

We now estimate the time $\tau_0$. 
For a  strip with resistivity $\rho(T_c)=240 \,\mu\Omega\,$cm, 
$w=120$ nm, $d=4$ nm, 
$\Lambda=45\,\mu$m, and a bias current of the order $I_c$, the crossing  
time is roughly $\tau_0\sim 10$ ps and corresponding vortex speed is 12 km/s.
This time is too short  for any  significant transfer of the electronic excitation
energy   into the substrate and surrounding strip area. 
Indeed, the phonon escape time was estimated as 160 ps 
in a strip of thickness $d=20$ nm, 
whereas the electron-phonon relaxation time is about 17 ps.\cite{Sem} 
During the time $\tau_0$ quasiparticles diffuse away from the vortex path by a 
short distance $(D\tau_0)^{1/2}\approx 8$ nm 
as estimated from the electronic specific heat $C_e=2.2$ kJ/m$^3$K and the normal-state 
resistivity at 10 K.\cite{Yang}

Hence quasiparticles remain practically within the belt of volume 
$V_b=\ell wd$ along   the vortex path. The quasiparticle energy density   
within the belt  is $(8\pi\xi/e\ell)(H_c^2/8\pi)(I/I_c)$.  
Taking $\ell\approx 3 \xi$, we see that for $I>I^*\approx I_c/3$ 
such an energy is sufficient to turn the belt
normal causing a dark count in the photon detector. 
We call this process at high currents $I>I^*$ a ``hot'' vortex crossing. 

Therefore, we conclude that the superconducting strip with a bias current in the interval $I^*<I<I_c$ is unstable 
with respect to the transition  into the normal state, that can be triggered by a vortex overcoming the barrier.
Clearly,     photons can trigger such a 
transition as well.  
The photon efficiency increases as $I$ approaches 
$I_c$ and so does the rate of dark counts. 

In fact, the true critical current of a strip, below which the strip remains superconducting,  is $I^*$.
At currents below $I^*$, the superconducting state is stable, but remains resistive due to the presence 
of quasiparticles in  normal cores of vortices crossing the strip. 
In this scenario a single vortex crossing leaves the strip in the superconducting 
state and thus we call this process  
a ``cold'' vortex crossing. 

\section{Comparison with experimental data }

\begin{figure}[tb]
\includegraphics[width=\columnwidth]{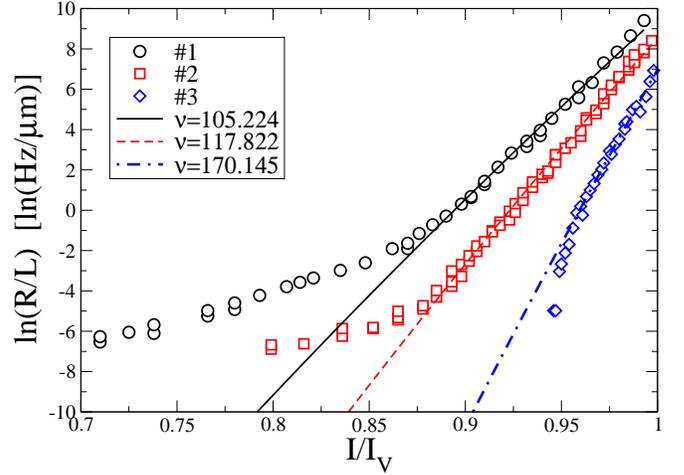}
\caption{(Color online) The dark count rates of three SNSPDs at 5.5 K by Bartolf et al.\cite{Bartolf} 
and fits based on Eqns.~(\ref{rate}) and (\ref{defY}).
The current is in units $I_V=\alpha I_c$, where $\alpha =0.72, 0.77, 0.60$ for samples 1, 2, and 3. 
$I_c$ is the critical current defined as the current at which the barrier for vortex crossings vanishes. 
At low currents electronic noise in the measurement setup dominates over vortex crossings. 
}
\label{fig2}
\end{figure}

In Fig.~\ref{fig2}  experimental   dark count rates are shown for three different NbN samples of SNSPDs.
\cite{Bartolf}  
We fit the data  using Eqs.~(\ref{rate}) and (\ref{defY}) by writing
\begin{eqnarray}
&&\ln(R/L)=\ln(a)+\ln[Y(\Phi_0I/\pi\nu cT)], \\
&&a=\frac{4Tc^2R_{\square}}{\Phi_0^2w}\left(\frac{\pi\nu^3}{2}\right)^{1/2}
\left(\frac{\pi\xi}{w}\right)^{\nu+1}.
\end{eqnarray}
The dimensions of samples 1, 2, and 3  are  $d=6$ nm,  $w=53.4, 82.9, 170.6$ nm, $L=73.9, 145.1, 141.4$ $\mu$m, respectively. 
The sheet resistance $R_{\square}=445, 393, 431\, \Omega$, and data were taken at $T=5.5$ K. 
According to Bartolf et al., at low currents the data was dominated by electronic noise in the measurement circuit.\cite{Bartolf} 
The data for samples 1 and 2 agree well with the theoretical results for high currents, while the data for sample 3 yield an unreasonably
large exponent $\nu$.

For sample 1 with fit parameter $\nu=\Phi_0^2/8\pi^2\Lambda$, we extract the Pearl length $\Lambda(5.5 {\rm K})=57.1$ $\mu$m,
and from  $\ln(a/L)$ we estimate  the coherence length $\xi(5.5 {\rm K})=3.9$ nm. The authors of Ref.~\onlinecite{Bartolf}
estimated  $\xi(0)=4$\,nm from independent measurements   of the upper critical field.  
They also estimated the  Pearl length for NbN films of thickness $d=6$ nm, 
 $\Lambda(0)=65.1$ $\mu$m,
 from known resistivity
$\rho_n$ and the 
superconducting gap $\Delta(0)\sim 2-3$ meV. \cite{Chockalingam2009}
By using Eq.~(\ref{Ic}), we find the critical current 
$I_c=20.1$ $\mu$A defined as the current at which the energy barrier vanishes for vortex crossings. 
The authors of Ref.~\onlinecite{Bartolf} defined the ``critical'' current $I_V=14.5$ $\mu$A using the 
1\% voltage criterion (current at which resistance is 1\% of the normal one). 
We see that the critical current defined through such a voltage criterion is less than the critical current
defined by the current at which the energy barrier vanishes, 
$I_V \approx 0.72\, I_c$. 
 
For the sample 2, we find the coherence length $\xi(5.5 {\rm K})=4.33$ nm  
and $\Lambda(5.5 {\rm K})=51\,\mu$m. Independent estimates given in Ref.~\onlinecite{Bartolf} are
$\xi(0)=4.2$ nm
and $\Lambda(0)=59.2\,\mu$m; the critical 
current $I_c=31.5\,\mu$A, while $I_V=0.77\, I_c$.\cite{Bartolf} 
We conclude that our model for vortex crossing rates describes satisfactory the dark count rates in   samples 1 and 2. 

Next we estimate  the peak  of the voltage pulse  for $I$ slightly below $I^*$:
\begin{equation}
V_{\rm peak}\approx \frac{c\Phi_0 \xi R_{\square}}{\pi e\Lambda w},
\end{equation}
and the duration of the pulse is $\tau_{\rm peak}<\Phi_0/cV_{\rm peak}$. 
For the sample 2 studied by Bartolf {\it et al.}\cite{Bartolf} we estimate 
$V_{\rm peak}\approx 0.8$ mV, 
while $\tau_{\rm peak}\approx 3$ ps slightly below $I^*$. 
For comparison, dark counts are characterized by peak voltages of  $\approx 1$ mV and by 
durations of several nanoseconds (FWHM $\sim 2.5$ ns \cite{Kit1}).
For dark counts the duration of pulses is caused by the current redistribution and thus
depends on the experimental setup used to detect the pulses. Note, that pulse duration differs significantly  from that 
caused by single vortex crossing without formation of normal belt.

The following experiment  could, in principle, distinguish between regimes at $I<I^*$ and at $I>I^*$:   
One induces a bias current in a thin-film ring   and measures the magnetic flux in the  ring as a function of time. 
For $I>I^*$, a single vortex crossing destroys superconductivity and the flux vanishes.  
The lifetime of this {\it persistent} current is $1/R$ and $R$ is determined by Eq.\,(\ref{rate}).
If $I<I^*$, the flux should decrease stepwise through multiple transitions  between quantized current states $I_n$, 
each transition corresponds to a single vortex crossing. 
In this case, the lifetime for the current $I_n$  is $1/R_n$ where $R_n$ 
is given by Eq.~(\ref{rate}) with $I=I_n$. 
The total decay time of the initial current $I_N$ will be $\tau=\sum_{n=1}^N R_n^{-1}$.
For 1D wires similar behavior due to phase slips  
was described by McCumber and 
Halperin. \cite{McCumber} 

In comparing theory and experiment, the issue of possible inhomogeneities of the thickness $d$ and the width $w$ is often raised. We note that the model developed here is only valid for  $w\ll \Lambda$. Each vortex in a narrow strip has mostly the kinetic energy of its supercurrents which are confined within an area of size   $\sim w\times w$. In other words, the model is not sensitive to inhomogeneities of $d$ and  of the edge roughness on scales small relative to $w$. 
  
Finally, it is worth mentioning that we assumed in this work that 
the strip temperature is equal to the bath temperature of the substrate. 
This may not always be the case in measurements of dark counts in photon detectors. 
After redistribution of the bias current, the normal belt induced by a  crossing  vortex  cools down.
The strip  can carry the superconducting current equal to the 
bias current $I$ only if the temperature drops below the value $T^*$ defined by the condition $I^*(T^*)=I$. 
Slightly below $T^*$ vortices can cross the strip inside the warmer belt whose temperature is 
close to $T^*$ or inside the cooler areas whose temperature is that of the bath. The rate of vortex 
crossings is determined by 
both processes and the latter dominates only in the limit of very large $L$. Again, we emphasize 
that the measured rate is 
higher than the calculated rate, and the difference is larger for small currents because for them $T^*$ is higher. 

\section{Conclusions}
In summary, we have found that the most plausible mechanism for dark counts in photon detectors 
is due to thermal fluctuations 
related to vortex crossings in the metastable current-carrying superconducting state,
which is realized at bias currents 
above some value $I^*\sim I_c/3$. 
We conclude by listing our main results:\\
(a) Vortices crossing the current-biased strip due to thermal fluctuations induce voltage pulses which 
can be detected experimentally. The barrier for vortex crossings vanishes at the critical current defined 
by Eq.~(\ref{Ic}). \\
(b) In narrow and thin strips, the superconducting state is unstable in the current interval $I^*<I<I_c$  and a transition  into the 
normal state is triggered by vortices crossing the strip accompanied by  
energy (heat) release.\\
(c) We estimated  the  threshold for ``hot'' 
vortex crossings  to be roughly $I^*\approx I_c/3$. \\
(d) Dark counts in current-biased superconducting 
strips reported in the literature
were observed in the regime of metastable superconducting state. \\
(e) At currents below $I^*$, vortex crossings do not induce transitions  into the normal state, 
but  still induce 
voltage pulses and the superconducting state is resistive due to the quasiparticles inside vortex cores 
of crossing vortices. We proposed a ring experiment, which allows to distinguish different decay processes
of circular currents above and below $I^*$.\\
(f) We estimated the amplitude and duration of ``cold'' voltage pulses which can be detected below $I^*$. 

Clearly it is desirable to test our theory by measuring I-V characteristics 
with a pulsed current technique to avoid heating. 
Further it will be interesting to study the rate and the shape of ``cold'' 
 pulses at currents below $I^*$ at different temperatures and see their evolution from thermally  
induced crossings to quantum tunneling. 

\acknowledgments
We are grateful to I. Martin, M. Rabin and D. Rosenberg for many useful discussions. 
Work at the Los Alamos National Laboratory was performed under the auspices of the
U.S.\ DOE contract No.~DE-AC52-06NA25396 through the LDRD program.
Work at the Ames Lab (VK) was supported by the DOE-Office of Basic
Energy Sciences, Division of Materials Sciences and Engineering under
Contract No.~DE-AC02-07CH11358.


\begin{thebibliography}{10}
\bibitem{BKT} In narrow thin-film samples with sizes $w\ll \Lambda$, the edge boundary 
conditions modify drastically the intervortex interactions, which are no longer logarithmic, 
see Ref.~\onlinecite{kogan2}. Therefore, the necessary condition for the BKT transition is violated. 
In our humble opinion, attempts to describe the physics of narrow thin-film superconducting samples with the help of 
the BKT theory are futile.
\bibitem{footnote1} 
The critical current $I_c$ defined as the critical current where the barrier vanishes doesn't describe any phase transition into the normal state, neither in 1D wires nor in narrow thin strips. As the current increases beyond $I_c$ in 1D wires
the normal state resistivity (Ohm's Law) is approached asymptotically by proliferation of phase slip centers.\cite{Tinkham}
\end{thebibliography}
\end{document}